\input{aipcheck}

\documentclass[
    ,final            
  ]
  {aipproc}

\layoutstyle{6x9}
\usepackage{natbib}
\newcommand{\sqdeg}{\,deg$^2$}

\begin{document}

\title{The ultracool field dwarfs luminosity function from the 
Canada-France Brown Dwarf Survey}


\author{C. Reyl\'e}{
  address={Observatoire de Besan\c{c}on, Institut Utinam, UMR CNRS 6213, 
   BP 1615, 25010 Besan\c{c}on Cedex, France}
}

\author{P. Delorme}{
  address={Laboratoire d'Astrophysique de Grenoble,Universit\'e
  J.Fourier, CNRS, UMR5571, Grenoble, France}
}

\author{X. Delfosse}{
  address={Laboratoire d'Astrophysique de Grenoble,Universit\'e
  J.Fourier, CNRS, UMR5571, Grenoble, France}
}

\author{T. Forveille}{
  address={Laboratoire d'Astrophysique de Grenoble,Universit\'e
  J.Fourier, CNRS, UMR5571, Grenoble, France}
}

\author{C. Willott}{
  address={University of Ottawa, Physics Department, 150 Louis Pasteur, 
   MacDonald Hall, Ottawa, ON K1N 6N5,  Canada}
}

\author{L. Albert}{
  address={Canada-France-Hawaii Telescope Corporation, 65-1238 Mamalahoa 
   Highway, Kamuela, HI96743, USA}
}

\author{E. Artigau}{
  address={Gemini Observatory Southern Operations Center c/o AURA, Casilla 603
   La Serena, Chile}
}

\begin{abstract}
The Canada-France Brown Dwarf Survey is a wide field survey for cool brown dwarfs conducted with the MegaCam camera on the CFHT telescope. Our objectives are to find ultracool
brown dwarfs and to constrain the field brown dwarf mass function from a large and homogeneous sample of L and T dwarfs. We identify candidates in CFHT/Megacam i' and z' images and
follow them up with pointed NIR imaging on several telescopes. Our survey has to date found ~50 T dwarfs candidates and ~170 L or late M dwarf candidates drawn from a larger sample of
1300 candidates with typical ultracool dwarfs i'-z' colours, found in 900 square degrees. We currently have completed the NIR follow-up on a large part of the survey for all candidates from
the latest T dwarfs known to the late L color range. This allows us to build on a complete and well defined sample of ultracool dwarfs to investigate the luminosity function of field L and
T dwarfs.\end{abstract}

\maketitle


\section{Introduction}

Thanks to recent and ongoing large scale surveys, hundreds of brown dwarfs have been discovered during the last decade. Still the luminosity function of field brown dwarfs is poorly constrained. Recently, \cite{Cruz.2007}
and \cite{Metchev.2008} obtained and used homogeneous sample to study the space density of
cool dwarfs (46 L dwarfs and 15 T dwarfs, respectively). 
In this work we attempt to compute the luminosity function for dwarfs from L5 to the latest T dwarfs.
We constructed a homogeneous and clean sample of 60 objects in this spectral type range, drawn from the Canada-France Brown Dwarf Survey (hereafter CFBDS).

We first briefly describe the survey and the observational strategy. Next we describe the sample including the determination of completeness and contamination, and we make a photometric classification. A preliminary luminosity function is given in the last section.

\section{Observations}
\label{obs}

The survey CFBDS is fully described in \cite{delorme.2008b}.
The $i'$ and $z'$ imaging part of the 900 \sqdeg\  Canada-France Brown
Dwarfs Survey is nearing completion to typical limit of $z'=22.5$.
The reddest sources are then followed-up with pointed J-band imaging at several observatories (La Silla, McDonald, Kitt Peak, La Palma) to distinguish brown dwarfs from other astronomical sources
and spectra are obtained for the latest type dwarfs. So far we have obtained $J$-band photometry for 670 objects, that is half of the ultracool candidates
selected on their $i'-z'>1.7$ colour and also spectra for 20 candidates (see Albert et al., in this volume).

Throughout this paper, we use Vega magnitudes for the $J$-band whereas AB magnitudes are used for the $i'$ and $z'$ optical bands.


\section{The sample}
\label{sample}

The CFBDS is composed of several patches -- contiguous areas on the sky -- with size ranging from 10 to 94 \sqdeg. As the priority was given for the reddest candidates for  $J$-band follow-up, we can now built a complete sample of late L and T dwarf candidates over 16 patches with total area of 520 \sqdeg. The sample contain all candidates with $z'<22.5$ and $i'-z'>2.0$ detected in this area.

\subsection{Photometric classification}

\begin{figure}[h]
\includegraphics[scale=0.8,scale=0.45,angle=-90,clip=]{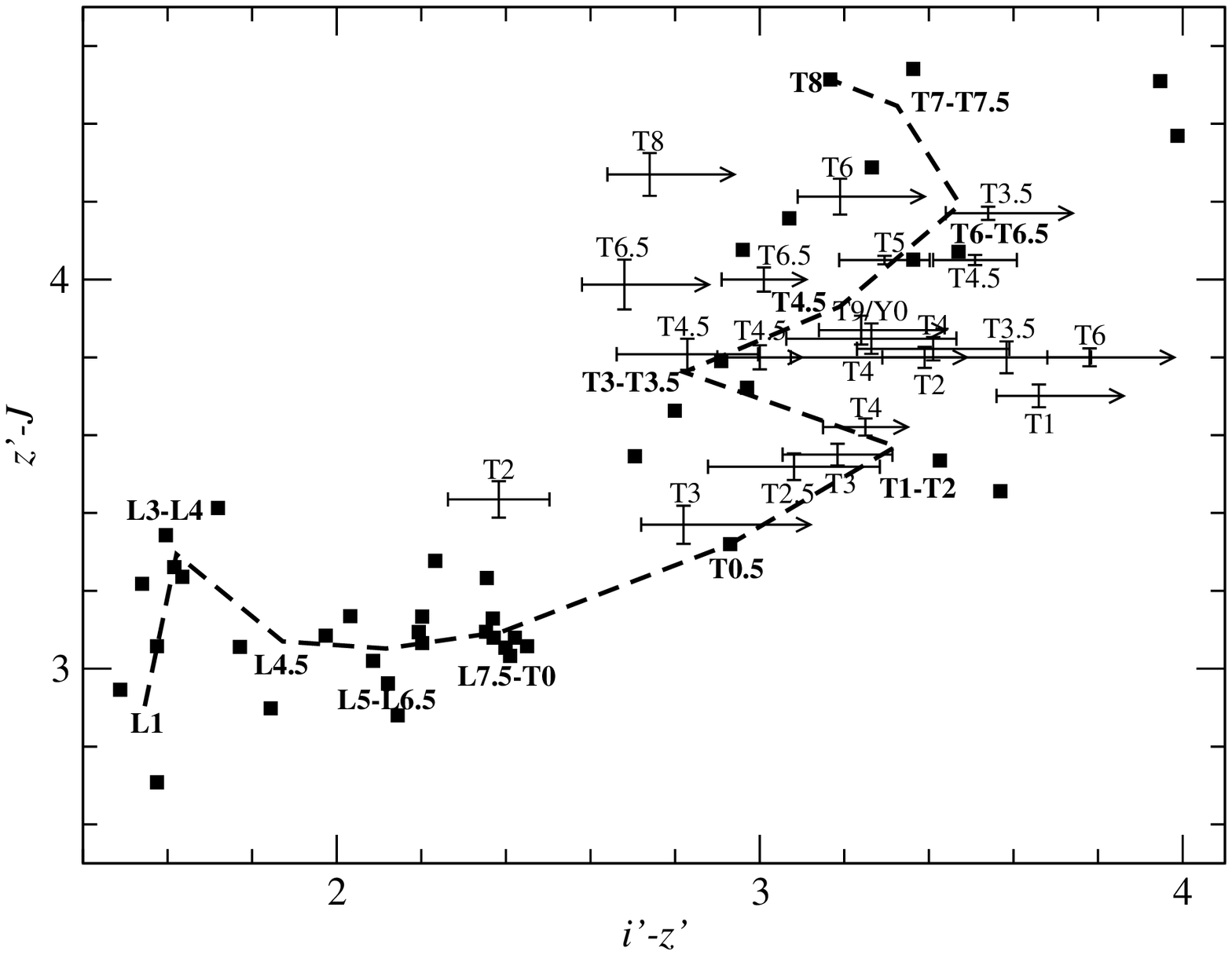}
\includegraphics[scale=0.8,scale=0.45,angle=-90,clip=]{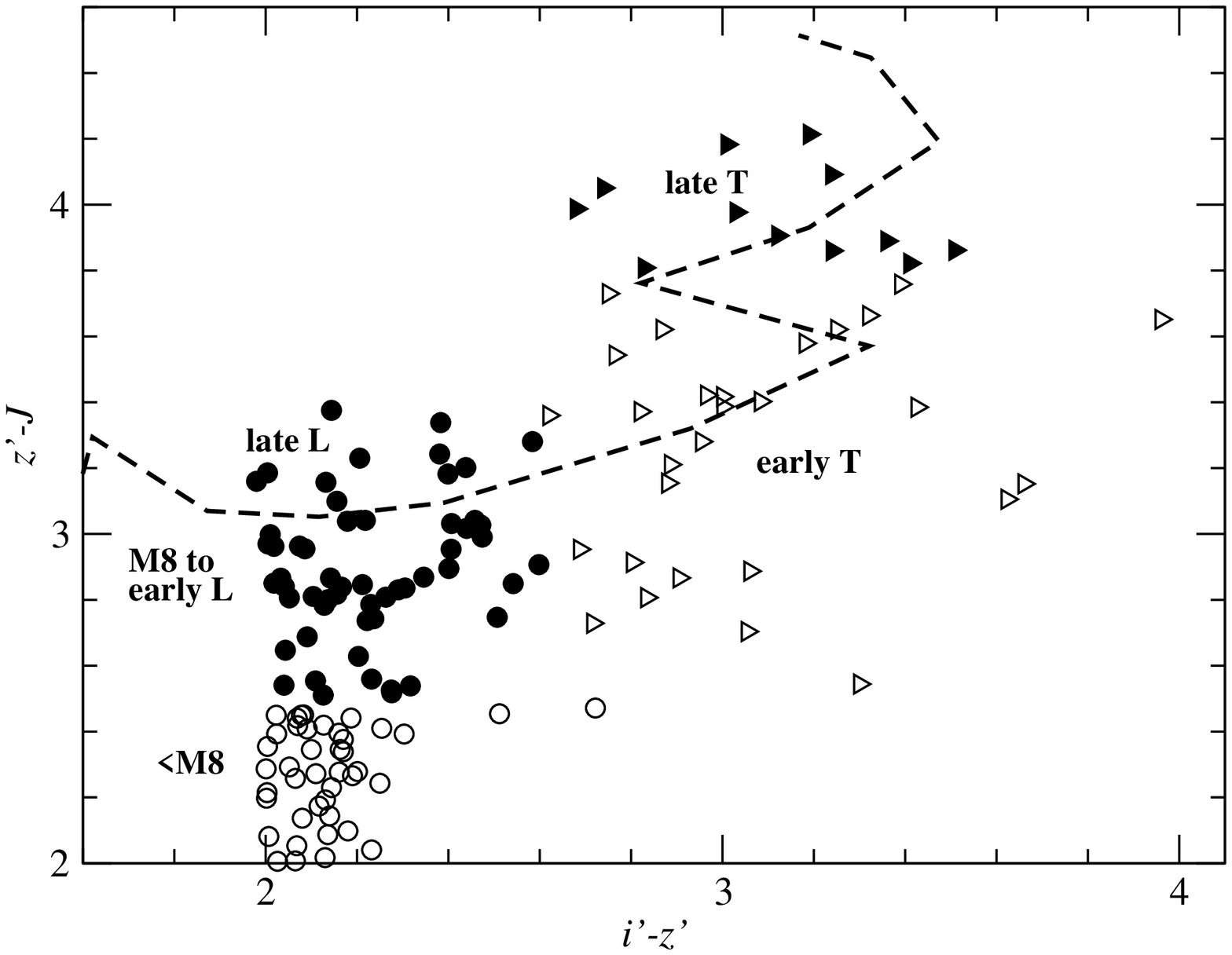}
\caption{Left: $z'-J$ synthetic colour versus $i'-z'$ synthetic colour computed from available spectra in the literature. 
The filled squares show the colours of each brown dwarf. The dashed line shows the mean colour-colour relation. Spectral type are given along this track, indicating the averaged colour over the brown dwarf with that spectral type. The open squares show the T dwarf candidates for which we obtained spectroscopic observations. An arrow indicates no detection in the $i'$-band.
Right: $z'-J$ versus $i'-z'$ diagram of the brown dwarf candidates in our sample with complete $J$-band follow-up. Open circles: dwarfs earlier than M8. Filled circles: M8 to L dwarfs. Open triangles: early T dwarfs. Filled triangles: late T dwarfs. The dotted line is the same as on the left panel.}
\label{iz-zJ_sample}
\end{figure}

This sample contains 191 objects that are candidates cooler than L5 on the basis of their $i'-z'$ colour. However, due to the photometric errors, many contaminants with true colour $i'-z'<2.0$ enter in this sample, such as the vastly more numerous M and early L dwarfs. A classification of the candidates is made on the basis of their position in the $i'-z'$/$z'-J$ diagram, as shown in Fig.~\ref{iz-zJ_sample} on the right panel. This classification is done using publicly available spectra from the L and T dwarf data archive\footnote{http://www.jach.hawaii.edu/~skl/LTdata.html}
of over 43 brown dwarfs with spectral types L1 to T8 on the \cite{Burgasser.2006} spectral type scale.
Thus synthetic colours in the Megacam and $J$ filters and are computed using detector quantum efficiency and transmission curves for the atmosphere, telescope, camera optics, and filters. 
The left panel on Fig.~\ref{iz-zJ_sample} displays the resulting colour-colour diagram.

Based upon this classification, 81 over the 191 objects remain L5 and later dwarf candidates, the others being dwarfs earlier than M8, artifacts or quasars. Still, M8 and early L dwarfs contaminate the sample as their $z'-J$ colours are similar to those of late L dwarfs.

\subsection{Completeness and contamination}

In order to estimate the completness of the sample, 1 500 000 fake cool
dwarfs built on the observed local PSF are added to the original frames. The resulting images go
through the analysis and selection pipelines used to select true candidates.
Thus we can count the number of detected objects as a function of magnitude on the detection image ($z'$-band) and colour. The completeness is given by the fraction of recovered fake objects at the end of the analysis process. The completeness is computed separately for each patch of the CFBDS.
 Moreover, the
measured magnitudes compared to the injected magnitudes allows us to derive the error distribution
probability that is used when computing the contamination of the sample.

As shown in Fig.~\ref{iz-zJ_sample}, our sample selects L5 and
cooler dwarfs. The earlier ones have $i'-z'<2.0$ and have no $J-$band follow-up observations.
However, due to photometric errors, part of these $<$L5 dwarfs are spread within the $i'-
z'>2.0$ sample.
All objects with $z'-J<2.5$ are removed from the sample. They are artifacts with obviously no $J$-band detection, quasars and dwarfs with true $i'-z'$ colour lower than 1.3, corresponding to spectral type earlier than M8.
However, ultracool dwarfs with true colour $1.3<i'-z'<2.0$ can not be discriminated this way because they have the same $z'-J$ colour as late L and early T dwarfs.

The estimation of the number of contaminants with true colour $1.3<i'-z'<2.0$ is done in two steps as explained below. This is done for each patch separately.
(i) Six patches of the CFBDS have $J$-band follow-up for all the retained candidates (that is with $i'-z'>1.7$). These patches contain 228 candidates over a total area of 199\sqdeg. The $z'-J$ diagnostic show that 65\% of these objects in the range $1.7<i'-z'<2.0$ are indeed $<$M8 dwarfs, artifacts or quasars. We consider that this value is the same for all patches.
(ii) We draw a photometric error in the $i'$ and $z'$ bands for the 35\% remaining objects. Objects with resulting $i'-z'>2.0$ and $z'<22.5$ are the contaminants. 

In this process, the contamination by objects with true colour $1.3<i'-z'<1.7$ is taken into account in an indirect way. Indeed, the observed sample with $1.7<i'-z'<2.0$ contains objects with true colour $1.3<i'-z'<1.7$.  Hence adding an error to the $1.7<i'-z'<2.0$ sample makes objects with true colour $1.3<i'-z'<1.7$ enter into the $i'-z'>2.0$ sample. We obtain 21 contaminating objects.

\section{Luminosity function}
\label{lf}

\begin{figure}[!h]
\includegraphics[scale=0.3,angle=0,clip=]{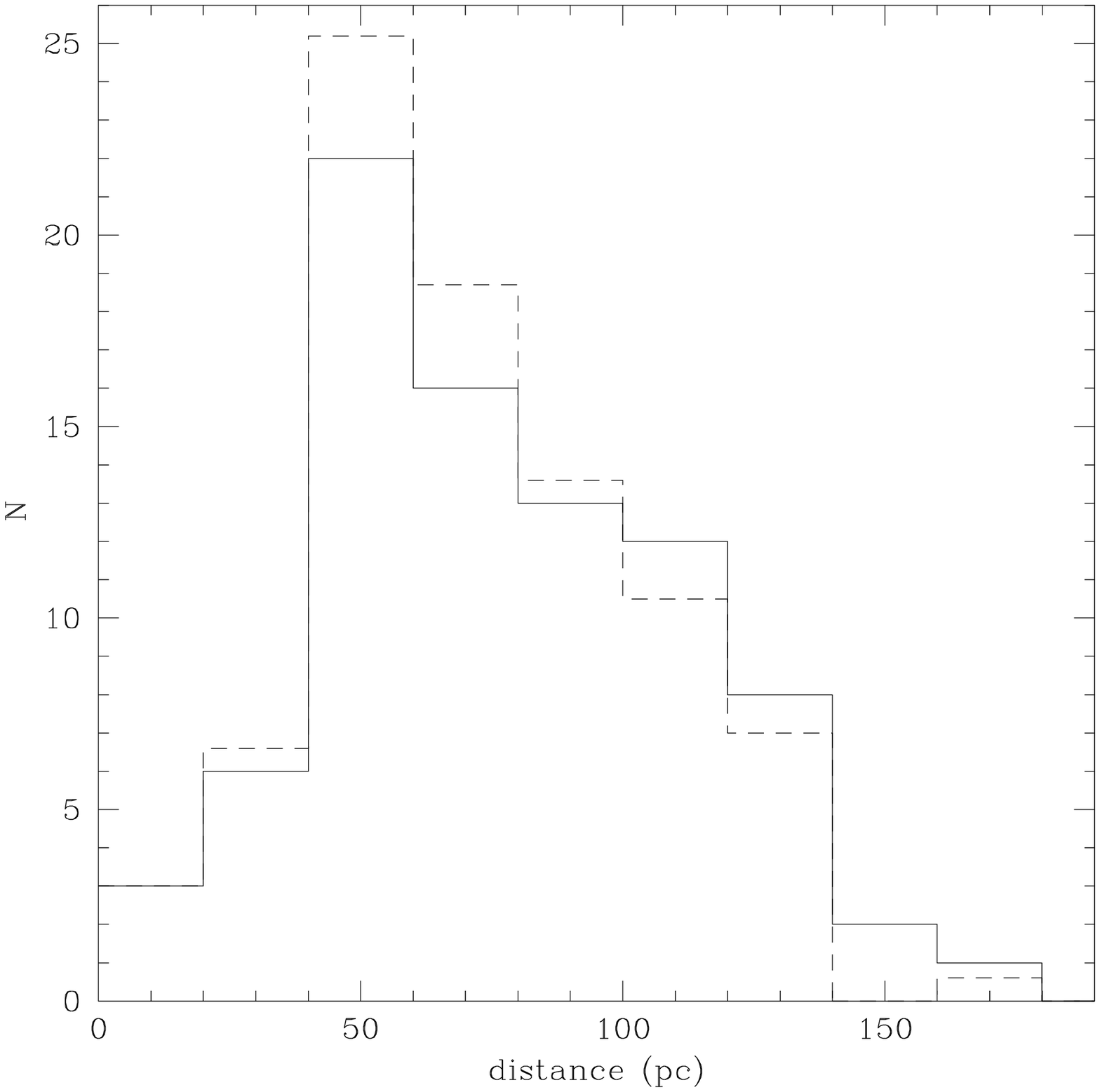}
\includegraphics[scale=0.3,angle=0,clip=]{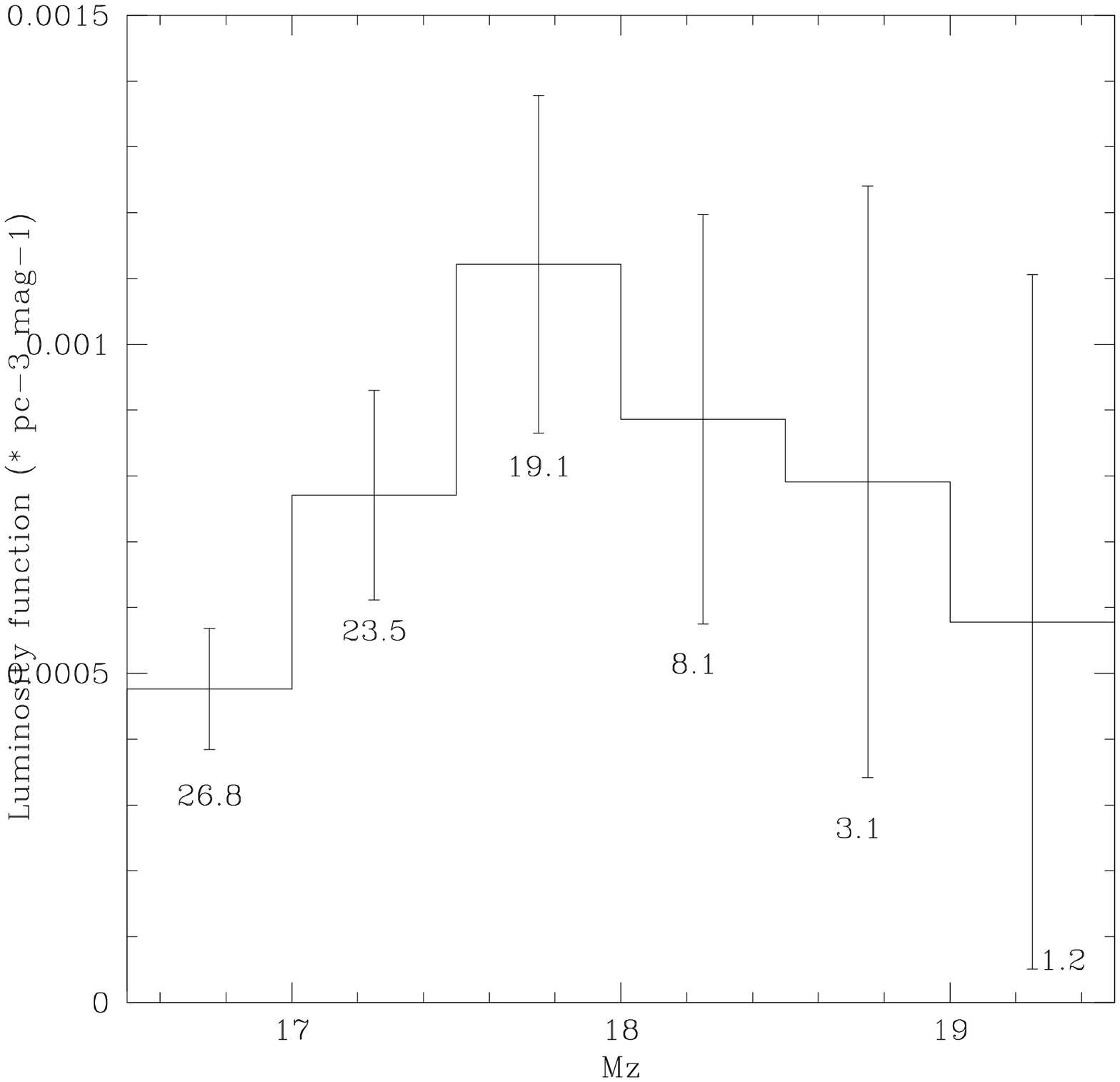}
\caption{
Left: Photometric distance distribution computed in the $z'$ band of our sample,
without (solid line) and with (dotted line) contamination and completeness correction.
Right: Luminosity function in the $z'$ band. The number of objects in
magnitude bins are also indicated, taking into
account completness and contamination
\label{dist}
}
\end{figure}

Among the publicly available spectra found in the L and T dwarf data archive, 32 have also measured parallax. They allow us to derive absolute magnitude-colour relations. It appears that the $i'-z'$ colour is a good luminosity estimator for late M and L dwarfs 
whereas  $z'-J$ is better for late T dwarfs.
These relations are used to derive photometric distances and a preliminary luminosity function in the $z'$ band, shown in Fig.~\ref{dist}. The absolute magnitude of early T dwarfs being nearly constant, this causes a bump in the luminosity function around $Mz'=17.5$. This corresponds to roughly $M_J=14$ where such a bump is already known to exist.

\bibliographystyle{aipproc}   

\bibliography{bib}

\IfFileExists{\jobname.bbl}{}
 {\typeout{}
  \typeout{******************************************}
  \typeout{** Please run "bibtex \jobname" to optain}
  \typeout{** the bibliography and then re-run LaTeX}
  \typeout{** twice to fix the references!}
  \typeout{******************************************}
  \typeout{}
 }

\end{document}